# Observation of pseudo-two-dimensional electron transport in the rock salt-type topological semimetal LaBi

Nitesh Kumar[1], Chandra Shekhar[1,*], Shu-Chun Wu[1], Inge Leermakers[2], Olga Young[2], Uli Zeitler[2], Binghai Yan[1], and Claudia Felser[1*]

[1]Max Planck Institute for Chemical Physics of Solids, 01187 Dresden, Germany
[2]High Field Magnet Laboratory (HFML-EMFL), Radboud University, Toernooiveld 7, 6525 ED Nijmegen, the Netherlands

Topological insulators are characterized by an inverted band structure in the bulk and metallic surface states on the surface. In LaBi, a semimetal with a band inversion equivalent to a topological insulator, we observe surface-state-like behavior in the magnetoresistance. The electrons responsible for this pseudo-two-dimensional transport, however, originate from the bulk states rather topological surface states, which is witnessed by the angle-dependent quantum oscillations of the magnetoresistance and *ab initio* calculations. As a consequence, the magnetoresistance exhibits strong anisotropy with large amplitude (~ $10^5$ %).

Topological insulators (TIs) are characterized by their conducting surface states due to the nontrivial topology of the bulk band structure.[1, 2] A three dimensional (3D) topological insulator has topologically protected conducting surface states in the form of an odd number of Dirac cones. Not limited to insulators, topological states have also been observed in semimetals for example Weyl semimetals and Dirac semimetals, in which conduction and valence bands disperse linearly through nodal points in all directions in the three-dimensional space.[3-7] When these bands are doubly degenerate, the system is called a Dirac semimetal. If the degeneracy is lifted by breaking time-reversal and/or inversion symmetry, properties of a Weyl semimetal can be observed. Weyl semimetals exhibit exotic Fermi arcs in the surface states[3] and interesting transport phenomena such as extremely large magnetoresistance (MR) and high charge carrier mobility.[8, 9] A group of topological materials exist, however, with a zero energy gap, which are actually semimetallic and referred as topological semimetals (TSMs). For example, the Heusler TIs usually exhibit an inverted band structure and the gapless feature at the Fermi energy.[10-12] Large MR and high mobility were also reported in these compounds such as LaPtBi.[13, 14] However, it is still illusive to distinguish the contributions by the bulk and surface states to the transport properties because of the semimetallic nature of TSMs.

Recently a new series of compounds, lanthanum monopnictides, were theoretically predicted to be TSMs[15], which stimulates the interest in their transport properties.[16] The last member, LaBi, is particularly interesting because of the largest spin-orbit coupling which may result into topological surface states. Our experiments reveal a large, unsaturated MR in LaBi, which is due to the electron-hole compensation in this semimetallic material. We observe a remarkable transverse MR of the order of $10^5$ % along with strong anisotropy which brings it in the line of the best known materials like NbP, WTe$_2$ and NbSb$_2$.[8, 17, 18] Strong Shubnikov-de Haas quantum oscillations develop at low temperatures which enables us to reconstruct the Fermi surface topology. The Fermi surface topology extracted from angular dependent magneto-transport experiments agrees well with our *ab initio* calculations. We observe an interesting strongly anisotropic Fermi surface that originates from the bulk electron pockets, rather than real surface states.

High purity La and Bi metals were weighed and transferred into an alumina crucible according to the composition La$_{0.33}$Bi$_{0.67}$ (15 g) inside argon filled glove box. The crucible was sealed in a quartz vessel under 3 mbar Ar pressure pressure to avoid the La attack on the quartz tube. The contents were heated at 1250 $^o$C with a heating rate of 100 $^o$C/h. This temperature was maintained for 1h followed by cooling to 1200 °C with a rate of 100 °C/h. After this the content was slowly cooled (1 °C/h) until 1050 °C for crystal growth. At this temperature the extra Bi flux was decanted out and the content was rapidly cooled down to room temperature. Cubic crystals of LaBi were retrieved and stored inside a glove box. The quality of the crystals was checked by using single crystal X-ray diffraction Bruker D8 VEsdNTURE X-ray diffractometer with $M_o$-$K_\alpha$ radiation and a bent graphite monochromator.

The transport measurements were performed using the ACT rotator option of physical property measurement system (PPMS, Quantum Design) with a maximum field of 9 T. The 35 T static magnetic field measurements were performed at the High Field Magnet Laboratory HFML-RU/FOM, member of the European Magnetic Field Laboratore (EMFL), in Nijmegen. Linear electrical contacts were made by 25 μm Pt or 40 μm Au wires using Ag epoxy as glue.

The Vienna *Ab-initio* Simulation Package (VASP) was adopted to perform the density-functional theory (DFT) calculations.[19, 20] The hybrid functional (HSE06)[21] with spin-orbital coupling was used to calculate the electronic structures. The bulk Fermi surfaces and the band structures were interpolated by maximally localized Wannier functions (MLWFs)[22-24]. The experimental lattice constant *a* = 6.5797 Å of rock salt LaBi was adopted.

LaBi has the largest spin-orbit coupling of all lanthanum monopnictides and it readily grows with Bi as a self-flux in contrast to e.g. LaSb where tin is used as flux and creates a possibility of elementary tin inclusion in the crystals. Moreover, LaBi is more stable than LaSb in air and moisture. It crystallizes in rock salt structure with space group *Fm*-3*m* wherein La and Bi atoms are arranged alternatively in all the three directions. The crystals obtained from Bi-flux growth are well faceted cube shaped with smooth (100) faces and can easily be cleaved along these faces.

We measured the electrical transport of our well oriented crystals (as confirmed from Laue diffraction) at temperatures down to 350 mK and magnetic fields up to 35 T. The current was applied along [100] and the field was rotated along different directions of the crystals. In order to improve statistical significance we have reproduced the measurements on many crystals. At zero magnetic field, the resistivity ρ shows purely metallic behavior, decreases linearly with temperature until 65 K below which it shows Fermi liquid behavior with $\rho_0 + aT^2$ relation (supplementary Fig. S1). The resistivity values are 4.95 x $10^{-5}$ Ω-cm at 300 K and 1.46 x $10^{-7}$ Ω-cm at 2 K resulting a residual resistivity ratio (RRR) = 339 confirming the high crystal quality, in fact the residual resistivity of LaBi is even less than in other known semimetals ($WTe_2$, NbP). Moreover, the conductivity per unit charge carrier at 300 K is of the same order as in highly metallic elements like copper.

When a magnetic field is applied, a metal to insulator-like transition is observed at low temperature and the transition temperature increases with increasing magnetic field (Fig. 1(d)). This was first observed in graphite but now it is well known for many semimetals possessing high mobility.[8, 17, 18, 25, 26] This phenomenon has broadly been attributed to the formation of excitons resulting in opening up of an excitonic gap in the material, which is enhanced in the magnetic field.[25] Semimetals with a smaller carrier density resulting in ineffective Coulomb screening and high mobility are ideal candidates to show metal to insulator-like transition. This behavior can also be understood by scaling the temperature dependent resistivity curves at different fields according to Kohler's rule as observed in $WTe_2$.[27] This behavior is illustrated by plotting the MR versus $\mu_0 H/\rho_0$ for different temperatures (Fig. S2 in the supplementary info). They all merge into the single behavior, which fit to the equation:

$$MR = A\left(\mu_0 H / \rho_0\right)^m$$

where *A* is a proportionality constant and *m* is the exponent. From the best fitting, *m* = 1.6 (*m*=2 corresponds to the perfectly compensated electron and hole system) which points out the compensated nature of LaBi. Moreover, the normalized temperature dependent MR values at different fields fall on top of each other. Hence, the low temperature phase is a metallic phase rather than an insulating one in high magnetic field. Transport at low temperature is dominated by electron charge carriers with a crossover temperature between 10 and 20 K (see Fig. S8) after which the hole charge carriers dominate.

Now we focus on magnetoresistance (MR) of LaBi, defined as *MR* (%) = 100 × ($\rho(B)$ − $\rho(0)$)/$\rho(0)$. This material exhibits a huge transverse MR at 2K and 9T (0.82 x $10^5$ %, see Fig. 2(b)) when the magnetic field is applied along [001]. Another crystal with lower RRR = 193 exhibited smaller magnetoresistance (0.38 x $10^5$ %) under the same conditions. The MR of both crystals increases quadratically with field. Extending the magnetoresistance measurements of LaBi to 35 T, we do not observe any sign of saturation (Fig. S9). The unsaturation behavior at high magnetic field has only been demonstrated in very few materials like NbP and WTe$_2$. The value of MR does not change significantly until 15 K after which it decreases sharply. Interestingly, we observe 100 % of further enhancement in the MR when the field is along [101] (MR = 1.5 x $10^5$ %, 2 K and 9 T) compared to [001] (MR = 0.8 x $10^5$ %, 2 K and 9 T) field direction. This anisotropic behavior in the MR is visualized in the polar plot of the resistivity where maxima and minima occur along [101] and [001], respectively (inset of Fig. 2(b)). Such a large MR anisotropy in LaBi is a clear indication of anisotropic Fermi surface character. In contrast, LaSb shows only 30 % increase in MR along [101] direction indicating smaller anisotropy compared to LaBi. *Ab initio* calculations[28] clearly show the larger anisotropy in LaBi compared to LaSb which will be discussed further in the following sections.

The field dependent resistivity shows SdH quantum oscillations, which are detected at low temperatures and fields as low as 2 T. To extract their amplitude of the oscillations, a third order polynomial was subtracted from the resistivity; the resulting oscillations at different temperatures are plotted against the inverse magnetic field (inset of Fig. 2(c)). These oscillations are highly periodic and their amplitudes diminish with increasing temperature beyond 10 K. We employed a fast Fourier transform (FFT) in order to extract the frequencies involved in the oscillations (Fig. 2(c)). We notice that two fundamental frequencies, $F_\alpha$ at 274 T from electron pocket, $\alpha$ and $F_\beta$ at 603 T from hole pocket, $\beta$ as well as second harmonic of $\alpha$-pocket at 547 T are observed when B is along [001]. The temperature dependence of the FFT amplitudes of the corresponding Fermi pockets is shown in Fig. 2(d), which follow the Lifshitz-Kosevich (LK) relation:

$$\frac{\Delta \rho_{xx}}{\rho_0} \propto e^{-2\pi^2 k_B T_D / \beta} \frac{2\pi^2 k_B T / \beta}{\sinh\left(2\pi^2 k_B T / \beta\right)}$$

where $k_B$ is Boltzmann's constant, and $T_D$ (Dingle temperature) and $\beta = ehB/2\pi m^*$ (with $m^*$ being the effective mass) are fitting parameters. Best fitting yields the values for $m^*$ and $T_D$ of $0.23 m_0$ and 8.9 K for the $\alpha$–pocket and $0.15 m_0$ and 9.1 K for the $\beta$–pocket, respectively, where $m_0$ is the bare electron mass. We analyze the most dominant frequency corresponding to the $\alpha$–pocket to obtain various parameters related to the Fermi surface, which play a significant role in electrical transport. Considering the circular cross section of the Fermi surface along [001], the Fermi area ($A_F$) that cover by electrons is found to be 0.026 Å$^{-2}$ from Onsager relation $F = (h/(4e\pi^2))A_F$ where $F$ is the frequency of oscillation. $A_F$ is used further to obtain the Fermi vector $k_F$ = 9.1 x10$^{-2}$ Å$^{-1}$. According to the relations $m^* = E_F/v_F^2$ and $v_F = \hbar k_F/m^*$ the values of Fermi energy ($E_F$) and Fermi velocity ($v_F$) are 2.76 meV and 4.6 x 10$^4$ m/s respectively.

Our *ab-initio* calculations clearly show the presence of electron and hole pockets located at the Fermi level. However, their presence can also be observed experimentally in a nonlinear Hall resistivity which otherwise should behave linearly. We observed nonlinear Hall effect, with a positive Hall constant at low fields which changes its sign at higher fields. On increasing the temperature, the Hall constant is positive in the entire field range. In order to quantitatively determine the carrier densities we calculate the conductivity tensor and apply a two-band model to separate carrier densities and their corresponding mobilities. The resultant hole and electron densities are 7.56 × 10$^{20}$ cm$^{-3}$ and 7.62 × 10$^{20}$ cm$^{-3}$ respectively, and the corresponding mobility values are 1.89 × 10$^4$ and 1.75 × 10$^4$ cm$^2$/Vs respectively, at 2 K. This is in accordance to the compensated nature of LaBi. High carrier concentration and large mobility explain the excellent conductivity in LaBi compared to other semimetals.

The band structure along high-symmetry lines for LaBi is shown in Fig. 3(a). The Fermi energy is 20 meV below the ideal electron-hole compensation point in the calculated band structure and crosses three doubly-degenerate bands (blue, green and red). The blue and green bands are hole pockets and the red bands are electron pockets. The band crossing points near *X*-points between La-5*d* and Bi-6*p* states present the band inversion indicating a 3D topological insulator. The exact position of the Fermi energy, $E_F$ (-20 meV) is determined by comparing the calculated angular dependence of the extremal cross-sectional area to the measured quantum oscillation frequencies. Calculations reveal two hole pockets ($\beta$ and $\gamma$) and three electron pockets ($\alpha$) in the first Brillouin zone (BZ). These electron pockets are identical in shape and they are located at every X-point. $\alpha$ and $\gamma$ are strongly anisotropic as shown in Fig. 3(b), but the pocket $\beta$ is nearly spherical and lies entirely inside $\gamma$ at the $\Gamma$-point. All pockets have only one extremal orbit when the field is applied along the crystallographic axes.

Experimentally, by tracking the angle dependent SdH frequencies it is possible to reconstruct the entire Fermi surface. In this process, we rotate the whole resistivity setup to track how the SdH oscillation frequency F changes as a function of the tilt angle with respect to the field in a broad range of more than 200°. The angle between the magnetic field and the current is varied in steps of 10° and the recorded resistivity and the corresponding FFT are shown in Fig. 4(a). We observe a shift of dominant $\alpha$ frequency as follows. At $\theta = 0°$ (B ∥ [001]), $F_\alpha$ is at 274 T (Fig. 4(b)) and on gradually tilting the field towards 90° (B ∥ [100]), $F_\alpha$ increases. When $\theta$ reaches 30°, a new frequency, $F_{\alpha'}$ is encountered just above $F_\alpha$ which decreases on increasing the angle further. This $F_{\alpha'}$ does not correspond to the fundamental frequency and is attributed to the neighboring (100) plane because of crystal symmetry. When B is applied perfectly along [100] at $\theta = 90°$, $F_{\alpha'}$ becomes the fundamental frequency corresponding to the (100) plane whereas $F_\alpha$ of the (001) plane disappears completely. This evolution of frequencies is exactly similar to the $\gamma$-band in simple cubic SmB$_6$ topological insulator.[29] From the full angular rotation, we tracked $F_\alpha$ and plotted it against $\theta$ (Fig. 4(b)) yielding intriguing results. $F_\alpha$ roughly follows the function $F_\alpha/\cos(\theta-n\pi/2)$ (shown by blue dashed lines) close to the principle crystallographic axes [001], [100] etc. as seen in Fig. 4(c). The similar inverse cosine behavior for $F_\alpha$ has also been observed in LaSb (from the same family of compounds) and the correlation has been argued to be arising from the topological 2D surface states. However, our *ab-initio* calculations in LaBi shows the presence of highly anisotropic 3D elongated ellipsoidal electron pockets (aspect ratio = 7.9). The cross section area of an ellipsoid and hence the corresponding frequency varies with the tilt angle as: $A \propto F_\alpha \propto \pi ab / \sqrt{\sin^2\theta + (a^2/b^2)\cos^2\theta}$, where $a$ and $b$ are semimajor and semiminor axes of the ellipsoid respectively. When a/b >> 1, the above relation reduces approximately to the inverse cosine for small $\theta$ values.[30] In LaBi, when the field is tilted further away from these axes, marked deviation from the inverse cosine relation is seen. We have mapped the experimental frequencies on the calculated angular dependence of Fermi surface cross sections (shown by red crosses in Fig. 4(c)) and observe a striking correlation between $F_\alpha$ and extremal cross section area of elongated 3D electron pockets. Hence, we consider the band corresponding to $\alpha$ to be only pseudo-2D in nature. The frequency corresponding to $\beta$ ($F_\beta$) is largely independent of the angle. This corresponds very well with the angular dependent extremal area of an almost spherical hole pocket ($\beta$) situated at the $\Gamma$-point of the Brillouin zone.

In summary, we show from electrical transport measurements and *ab-initio* calculations that despite having topologically protected insulating gap, LaBi behaves as a semimetal with very high bulk conductivity. At low temperature, the transport in lanthanum monopnictides is dominated by highly elongated 3D electron pockets centered at *X*-points in the Brillouin zone which behave like pseudo-2D and roughly follow the inverse cosine rule. Additionally, co-presence of electron and hole pockets makes LaBi a compensated material, with a huge unsaturated magnetoresistance and remarkable anisotropy.

This work was financially supported by the Deutsche Forschungsgemein- schaft DFG (Project No. EB 518/1-1 of DFG-SPP 1666 "Topological Insulators", and SFB 1143) and by the ERC (Advanced Grant No. 291472 Idea Heusler). The authors declare that they have no competing financial interests.

*Corresponding authors

shekhar@cpfs.mpg.de and felser@cpfs.mpg.de

**Figure captions**

FIG. 1. Schematic band structure, unit cell, photograph of the crystal and temperature dependent resistivity of LaBi. (a) Schematic band structure of an ideal topological insulator (left panel) where the topological gap and the corresponding surface states are at the Fermi energy. Schematic band structure of LaBi with the topological gap and the surface states below the Fermi energy (right panel) in addition to a trivial valence band crossing the Fermi energy. (b) Crystal structure of LaBi where La and Bi are represented with red and green spheres respectively. (c) Typical crystal of LaBi obtained from the flux growth technique where crystal edges are marked with red dashed lines. (d) Temperature dependent electrical resistivity of LaBi at 0 T and fields up to 9 T. The blue dashed line is a guide to the eye to track the metal to the insulator-like transition with increasing field.

FIG. 2. Magnetoresistance and SdH quantum oscillations of LaBi (a) Field dependent transverse resistance of LaBi at different temperatures starting from 1.85 K to 20 K with a maximum field of 9T. (b) Transverse magnetoresistance of LaBi at fields along [001] and [101] directions at 2 K. Inset of Figure 2b shows the polar plot of MR at 5T wherein maxima and minima occur at fields along [101] and [001] family of directions correspondingly. Inset of (c) shows SdH quantum oscillations after subtraction of third order polynomial plotted against the inverse of the magnetic field at temperatures from 1.85 K to 10 K. (c) Corresponding FFT amplitudes of SdH oscillations depicting fundamental frequencies $F_\alpha$ (274 T), $F_\beta$ (603 T) and the second harmonic of $F_\alpha$ (546 T). (d) Temperature dependent FFT

amplitudes of $F_\alpha$ (274 T) and $F_\beta$ (603 T) along with their fittings according to Lifshitz-Kosevich relation to obtain their corresponding effective masses.

FIG. 3. Band structure and the first Brillouin zone of LaBi (a) Bulk band structure of LaBi. The gray dash line is the Fermi energy 20 meV below the compensation point. A topological gap appears near X-point of the Brillouin zone. (b) The Fermi surface in the first Brillouin zone at Fermi energy. There are two hole pockets centered at the $\Gamma$-point ($H_1$ corresponds to $\beta$ band) and three identical electron pockets (E corresponds to $\alpha$ band) centered at X-points.

FIG. 4. Angular dependent SdH oscillations, corresponding FFTs and Fermiology of LaBi. (a) SdH oscillations amplitude (after polynomial (cubic) background subtraction) for tilt angles 30° to 150° (step size 10°). (b) Corresponding FFT amplitudes showing the angular dependence of $F_\alpha$ and $F_\beta$. The inset shows the scheme for the field rotation in the (010) plane. (c) Angular dependence of the SdH oscillation frequencies (solid circles) along with the calculated frequencies obtained from the extremal area of 3D Fermi surface cross sections (red cross for $\alpha$ electron pockets, green cross for $\beta$ hole pocket) and inverse cosine relation (blue dashed lines). The purple solid circle is the frequency obtained from high magnetic field measurement up to 35 T.

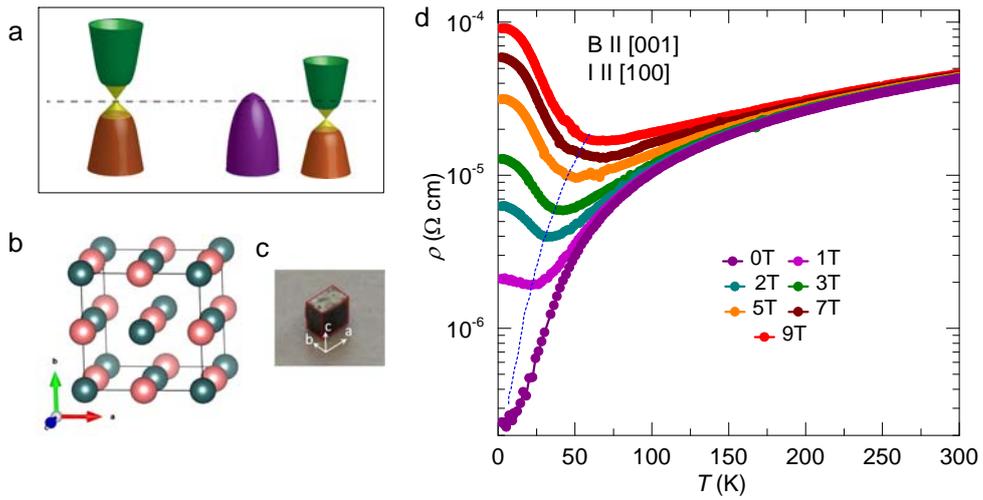

FIG. 1

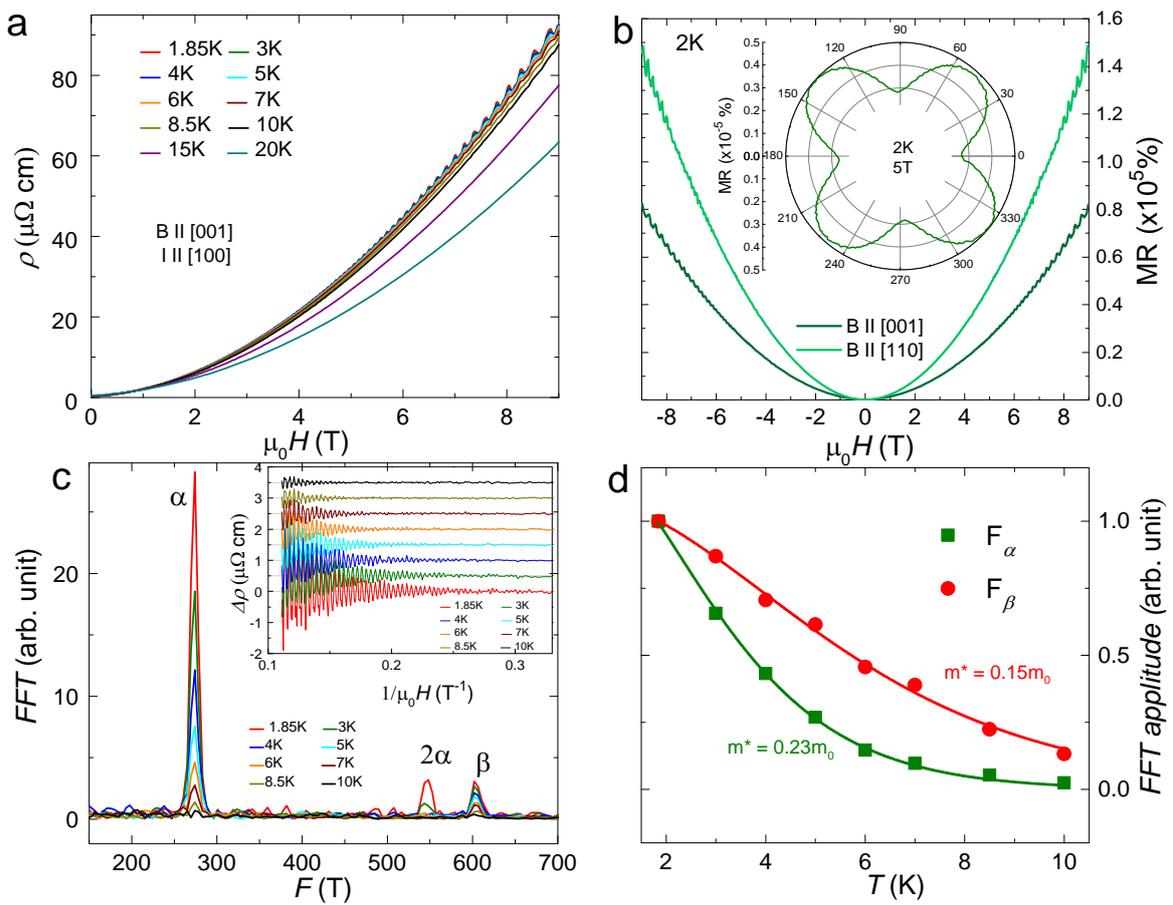

FIG. 2

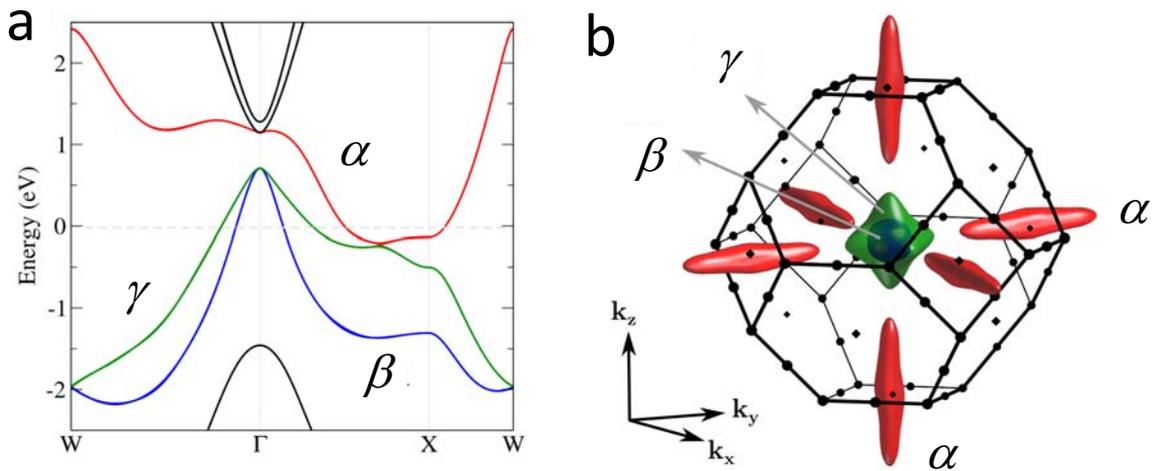

FIG. 3

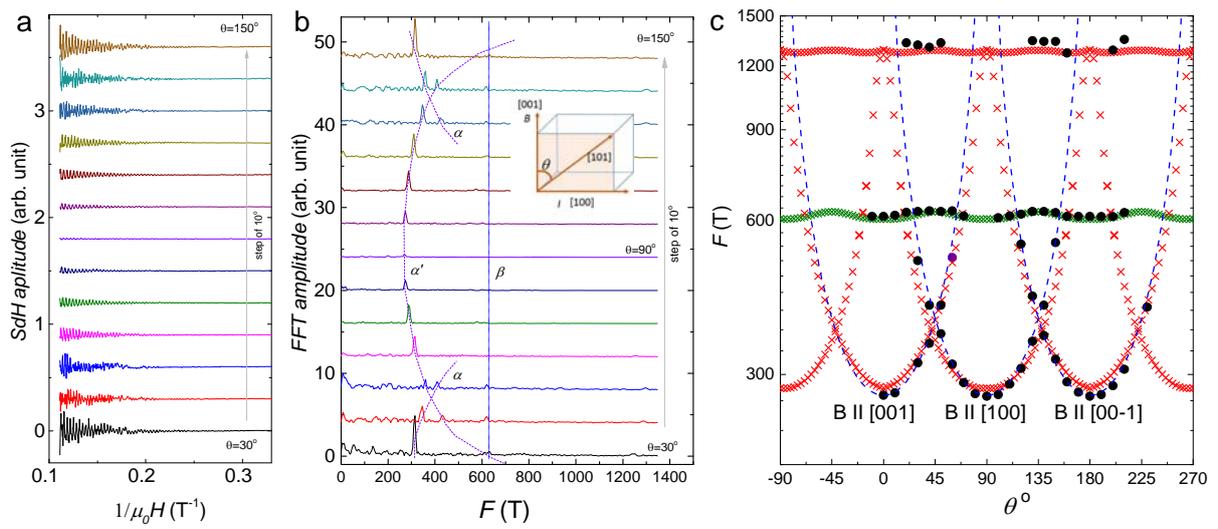

FIG. 4

# Supplementary Information

# Observation of pseudo-two-dimensional electron transport in the rock salt-type topological semimetal LaBi

Nitesh Kumar[1], Chandra Shekhar[1], Shu-Chun Wu[1], Inge Leermakers[2], Olga Young[2], Uli Zeitler[2], Binghai Yan[1], & Claudia Felser[1]

**Fermi liquid behavior**

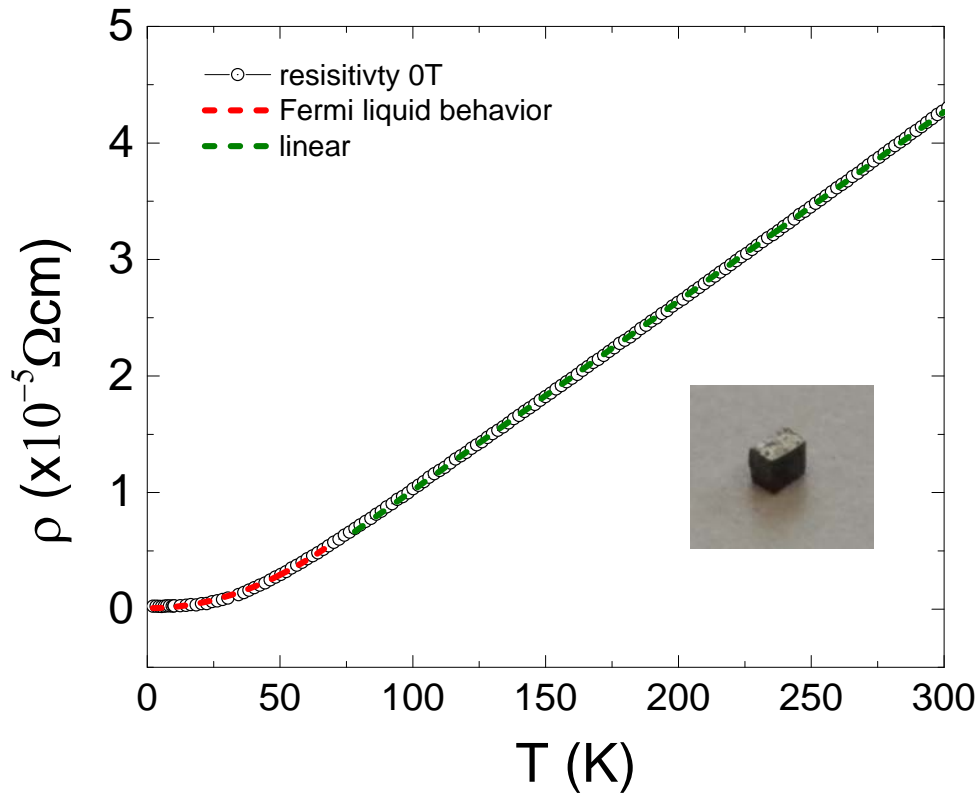

FIG. S1. Temperature dependent resistivity at zero field showing Fermi liquid behavior (fitted with red dotted line) in low temperature resistivity and linear behavior (fitted with dotted green line) at high temperature. The inset shows a typical crystal of LaBi obtained from flux growth technique.

## Scaling of temperature dependent resistivity by Kohler's rule

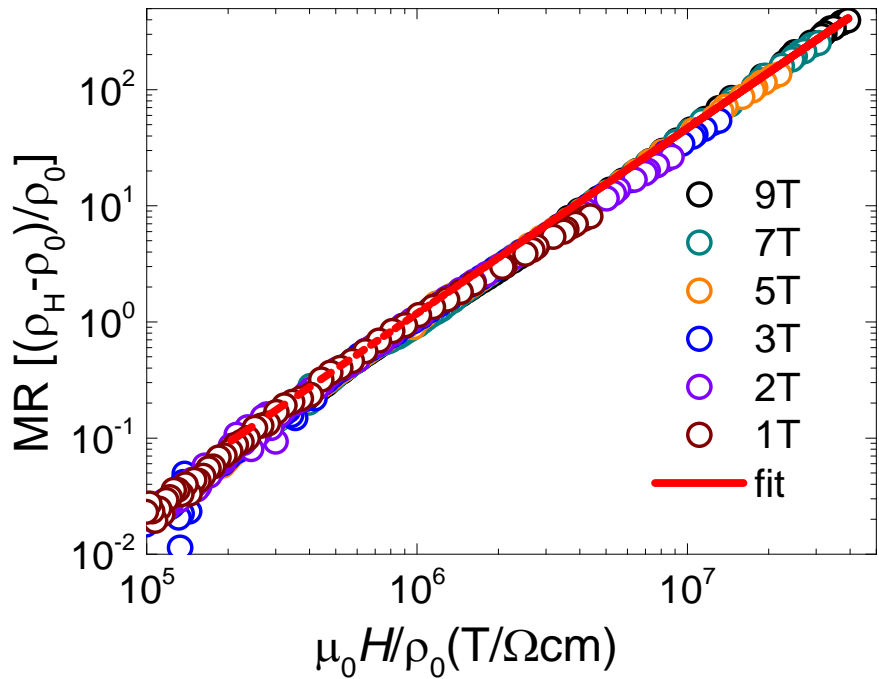

FIG. S2. Temperature dependent MR plotted against magnetic field over zero field resistivity. The overall behavior is fitted (res line) with the relation: $MR = A(\mu_0 H/\rho_0)^m$.

## Comparison of MRs at different fields

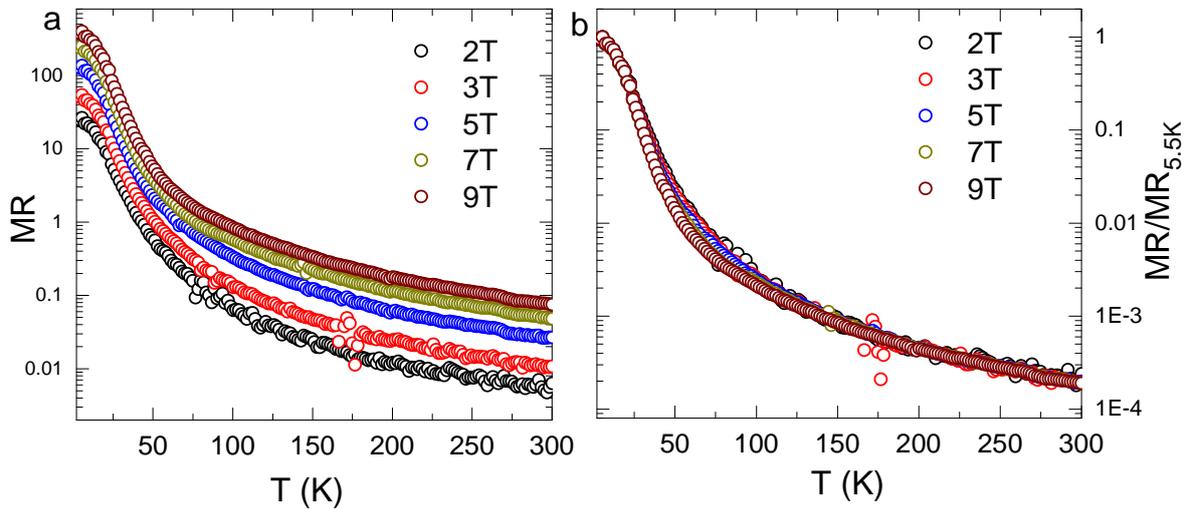

FIG. S3. (a) Temperature dependent MR at different magnetic fields. (b) Temperature dependent MR at different magnetic fields normalized by their corresponding MR values at 5.5 K.

**Angle dependent MR**

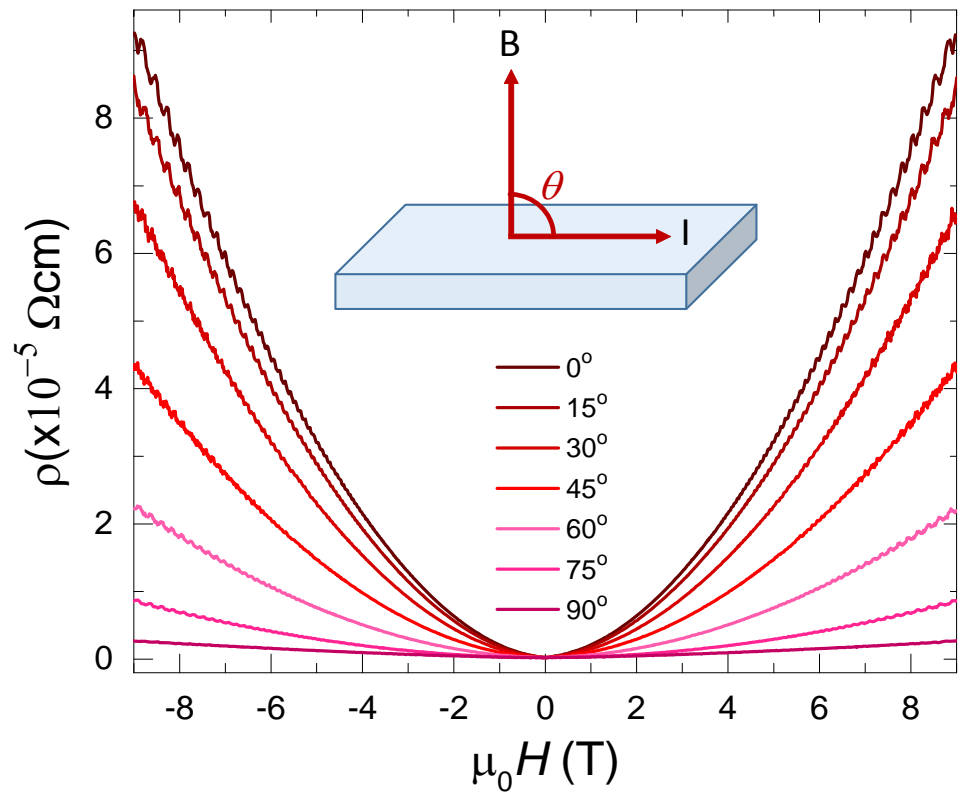

FIG. S4. Angle dependent MR of LaBi at 2K; inset shows the profile of crystal movement in the fixed field direction.

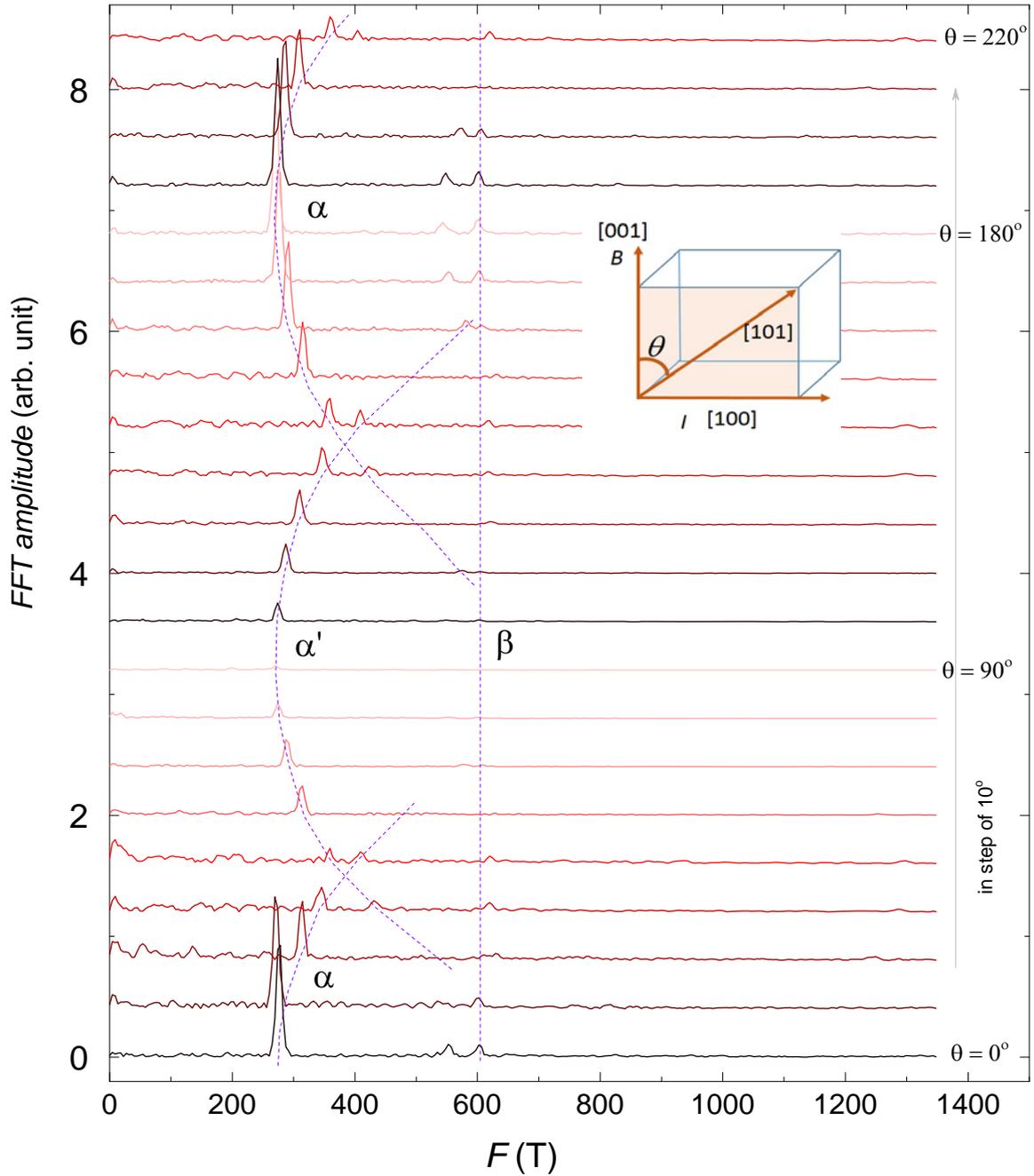

FIG. S5. Angle dependent FFT for field rotation in the angle range 0° to 220° in a step of 10°. Frequencies corresponding to $\alpha$, $\alpha'$ and $\beta$ are tracked by blue dashed lines. The inset shows the schematic of field rotation in the plane (010).

**Polar plot**

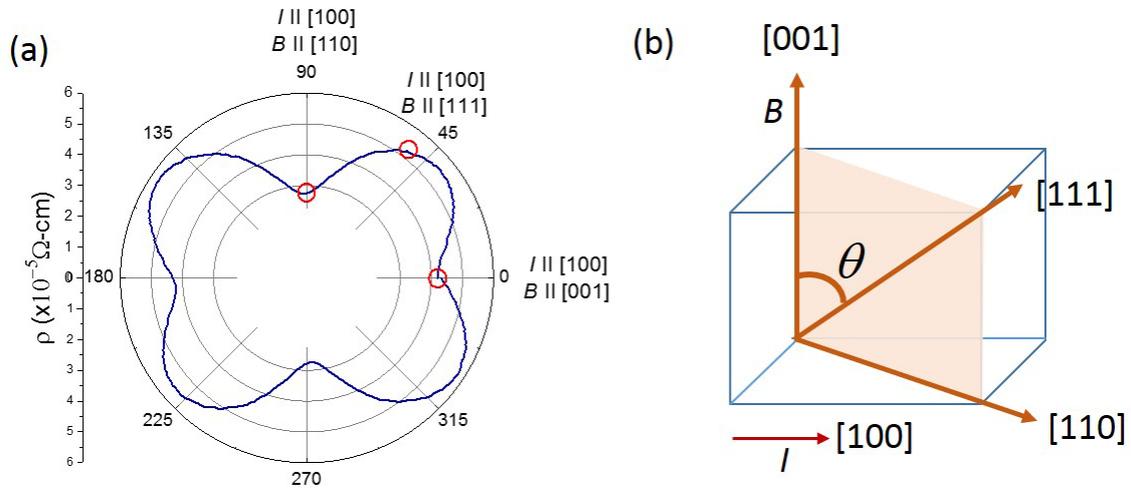

FIG. S6. (a) Polar plot of resistivity when the current is passed along [100] and the field is rotated in the direction [001]→[111]→[110]. (b) Schematic of the crystal orientation and its rotation with respect to the magnetic field.

**Temperature dependent MR**

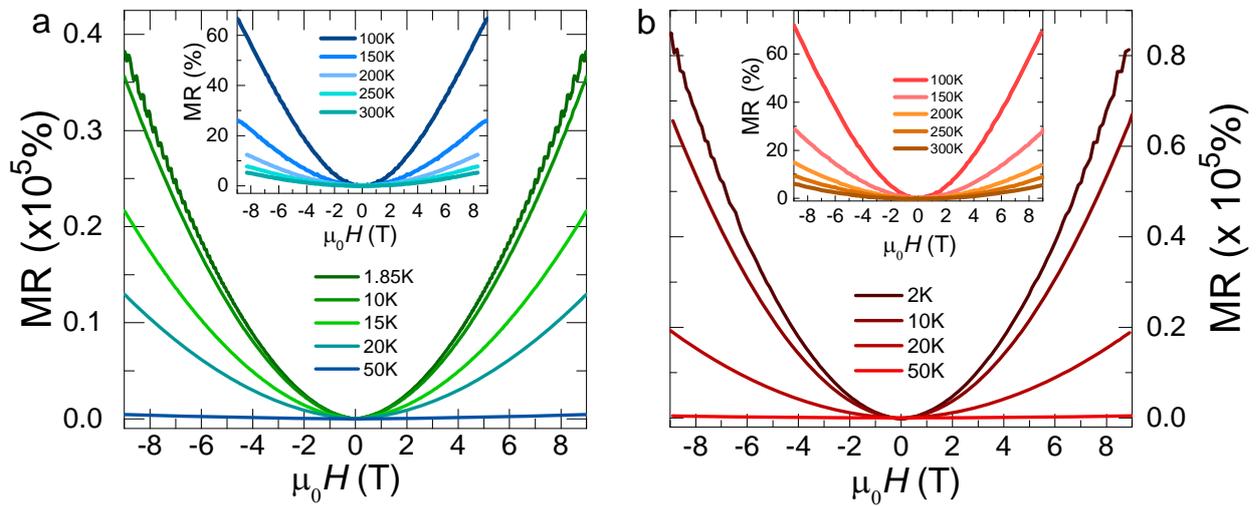

FIG. S7. MR of LaBi crystals at different temperatures for crystals (a) RRR = 193 and (b) R = 339.

**Two band model**

As the Fermi surface of LaBi has four fold symmetry, the Hall conductivity tensor can be defined as:

$$\sigma_{xy} = \frac{\rho_{yx}}{\rho_{yx}^2 + \rho_{xx}^2}$$

According to semiclassical Drude model individual electron and hole conductivities can be summed up to obtain overall transverse conductivity tensor as follows:

$$\sigma_{xy} = \left[ n_h \mu_h^2 \frac{1}{1+(\mu_h B)^2} - n_e \mu_e^2 \frac{1}{1+(\mu_e B)^2} \right] eB$$

$$\sigma_{xx}(B=0) = (n_h \mu_h + n_e \mu_e) e$$

Here, $n_h$, $\mu_h$ are density and mobility of hole whereas as $n_e$, $\mu_e$ are density and mobility of electron respectively; $e$ is the electronic charge. Using these equations, Hall conductivity can be fitted to obtain hole and electron density and mobility as fitting parameters.

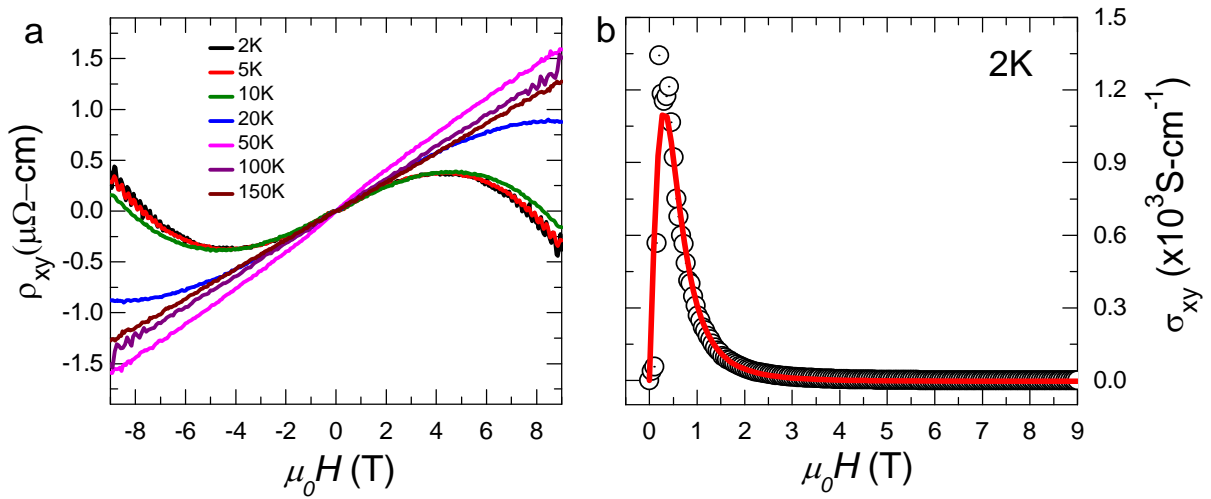

FIG. S8. (a) Hall resistivity of at different fields as a function of magnetic field. For temperature until 20 K, Hall resistivity is nonlinear after which it varies linearly. (b) Hall conductivity (black open circles) as a function of field at 2K along with the fitting (red solid line) according to the two band model.

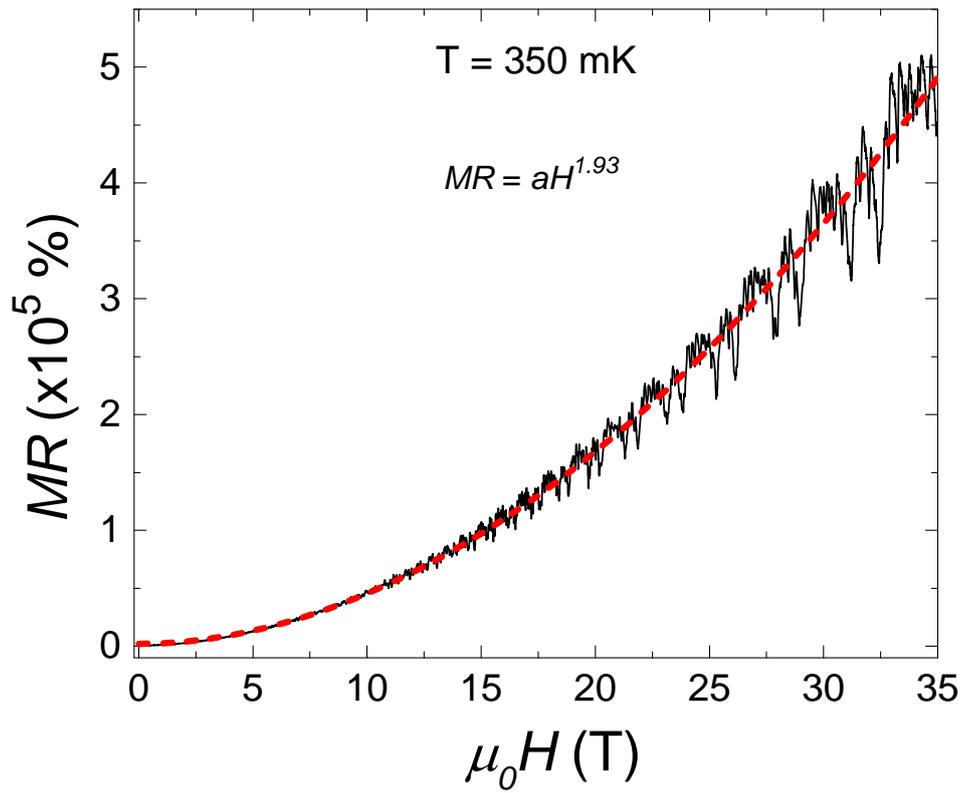

FIG. S9. High magnetic field MR with a maximum field of 35T. The MR remains unsaturated until 35T and behaves almost parabolic ($MR = aH^{1.93}$) with field as shown by the red dotted line.